\documentclass[preprint,12pt,authoryear]{elsarticle}

\usepackage{amsmath,amssymb}
\usepackage{float}
\usepackage{subcaption}
\usepackage{xcolor}
\usepackage{graphicx}
\usepackage{booktabs}
\usepackage{makecell}
\usepackage[table]{xcolor}
\usepackage{multirow}
\usepackage{url}
\usepackage{algorithm}

\journal{Renewable Energy}

\begin{document}

\begin{frontmatter}



\title{Multidisciplinary Design Optimization of Wave Energy Converter Farms Considering Uncertainty through Polynomial Chaos Expansion}

\author{\textbf{Kapil Khanal$^{*a}$, Nate DeGoede$^{*b}$, Maha N. Haji$^{b}$}}


\affiliation[a]{organization={Systems Engineering, Cornell University},
            city={Ithaca},
            state={NY},
            country={USA}}

\affiliation[b]{organization={Mechanical and Aerospace Engineering, Cornell University},
            city={Ithaca},
            state={NY},
            country={USA}}

\begin{abstract}
In this paper, a multidisciplinary design optimization problem under uncertainty is formulated for wave energy converter array. An array of heaving point absorbers for grid-scale energy production with decision variables and parameters chosen from the coupled disciplines of geometry, hydrodynamics, layout, and trajectory optimization thus resulting in a control co-design formulation of the plant and the control together. We study the benefits of MDO, and WEC farms and show that placing WEC farms in an array and optimizing it under uncertainty is significantly better than optimizing a WEC and placing it in an array. We vary the wave energy converter (WEC) dimensions, array layout, and control gain to minimize the power per volume. Uncertainty in the electrical power is handled using regression based on polynomial chaos expansion (PCE) method at each design iteration. Traditional WEC farm design optimization approaches often neglect the multidisciplinary, coupled nature of WECs and the inherent uncertainty in ocean wave conditions and control responses. This leads to designs that may under perform in real-world environments. In this work, we address this limitation by incorporating uncertainty directly into the design optimization process using the technique of polynomial chaos expansion (PCE) to quantify the variability of the performance due to uncertain wave environment. It is shown that the robustly designed WEC farms are significantly less sensitive to the input variations. 
\end{abstract}




\end{frontmatter}


\section{Introduction}
Wave energy converters (WECs) convert the oscillatory motion of ocean waves into usable mechanical or electrical energy~\citep{Budal1977}. For grid-scale deployment, multiple WECs must operate collectively, forming so-called wave farms comprised of arrays of WECs. According to \citet{Kilcher2021}, such farms could theoretically supply up to 34\% of U.S. electricity demand. Yet, large-scale development remains limited due to high capital costs, permitting challenges, and unresolved technical complexities~\citep{pacwave}.

A critical difficulty in designing WEC farms arises from hydrodynamic interactions among devices, identified by \citet{BABARIT201368} as the ``park effect," in which radiated and diffracted waves from one WEC influence the performance of others. These coupled effects mean that total farm power output depends on array geometry, device spacing, control strategy, and the local wave environment. Numerous studies have examined these factors from different perspectives. For example, \citet{BOZZI2017378} simulated WEC array configurations under real-sea conditions off the Italian coastline and showed that the dominant wave heading strongly affects optimal layouts; however, their study used brute-force optimization over only four candidate configurations. To reduce computational cost, \citet{giassi_layout_2018} applied cluster-based optimization to minimize the levelized cost of electricity (LCOE), while \citet{LYU2019106543} employed genetic algorithms to optimize the geometry and layout of three-, five-, and seven-body arrays. Their results demonstrated up to a 39\% increase in power production through joint optimization of design variables and further revealed that, under irregular waves with unconstrained impedance-matching and passive-derivative control strategies, the optimal array layout becomes asymmetric. Although these configurations achieved improved hydrodynamic performance, the resulting designs, featuring slightly different radii and drafts for each WEC, pose challenges for manufacturing and economies of scale. Building on these findings, subsequent studies~\citep{BABARIT201244, goteman2015geometries, ABDULKADIR2023113818, Coe2021} have reinforced the importance of simultaneous consideration of geometry, layout, and control, an approach broadly recognized as control co-design (CCD)~\citep{engineering_codesign}, to maximize array efficiency.




Despite these advances, most WEC farm optimization studies neglect uncertainty in ocean wave conditions and control responses. Parameters such as wave height, period, and heading vary stochastically, and deterministic designs often fail to perform reliably under real environmental variability. To achieve practical and economically viable solutions, optimization must therefore address the coupled influences of geometry, layout, power take-off (PTO) dynamics, and control design under uncertainty. In this work, we consider a unidirectional coupling between design and control, as defined by \citet{couplingCCD}, and formulate an optimization problem for a six-body WEC farm operating in irregular waves with uncertain wave direction and significant wave height.

Because WEC performance depends simultaneously on geometry, hydrodynamics, control, and cost models, the system is inherently high-dimensional and tightly coupled, making it well-suited to a multidisciplinary design optimization (MDO) framework~\citep{mdobook}. The MDO approach allows these interdependent disciplines to be optimized concurrently, producing more integrated and robust designs than traditional sequential methods~\cite{sobieszczanski-sobieski_sensitivity_1990}.

Uncertainty in ocean conditions exerts a strong influence on array performance and must be incorporated directly into the optimization process. Prior work has explored the effects of stochastic parameters such as wave incidence angle and spacing constraints on array efficiency. For instance, \citet{Sinha2016} examined the influence of wave incidence angle on power absorption across various fixed array configurations, showing that linear and grid-type layouts are more sensitive to wave direction than circular or concentric arrays. Similarly, \citet{sharp2018wave} investigated the role of minimum separation distance on array performance metrics, finding that variations in spatial constraints can significantly alter the optimal layout. While these studies provided valuable insight into the sensitivity of array performance to uncertain parameters, they generally relied on simplified hydrodynamic models (e.g., \citet{mcnatt2015novel} ) and did not integrate across hydrodynamics, control, and economic disciplines.

To address this gap, we integrate uncertainty quantification directly into the MDO process using the Polynomial Chaos Expansion (PCE) method. Uncertainty-based MDO has already demonstrated significant value in aerospace systems, where PCE models efficiently approximate stochastic behaviors with minimal sampling cost~\citep{Umdo_review}. More broadly, PCE-based surrogate modeling has proven effective across engineering disciplines for efficiently approximating stochastic system behavior with limited samples. In aerospace applications, least-squares PCE surrogates have enabled robust aerofoil shape optimization ~\citep{aerospace_uq}, while in wind energy they have been used to estimate annual energy production under uncertain wind conditions~\citep{wes-4-211-2019}. Similarly, \citet{mda_uncertainty} employed semi-intrusive PCE models to propagate modeling uncertainty in coupled MDO frameworks. More recently, PCE has been applied to marine systems to analyze the sensitivity of floating vessel wave-frequency responses, such as root-mean-square motions, to uncertain system variables~\citep{UQoffshore} such as center of gravity, vessel mass, roll moment of inertia, pitch moment of inertia etc to enable risk-neutral optimization. Similarly WEC control parameters in a wave farm, such as stiffness and damping, under uncertain wave directions~\citep{GAMBARINI2023112478} are to be considered. They optimized control strategies independently of geometry and array layout. This decoupled approach limits cross-disciplinary performance improvements that can be achieved through full CCD, a methodology shown to deliver substantial gains across complex engineered systems~\citep{GRASBERGER2024120234,garcia_co_design, engineering_codesign}.

To close this gap, this chapter introduces a robust CCD optimization framework for WEC farms that jointly tunes geometry, layout, and control parameters under uncertain sea states. Designs that neglect parameter uncertainty are often highly sensitive to variations in environmental conditions, resulting in suboptimal or unreliable performance. To address this, the proposed framework integrates a non-intrusive, regression-based PCE surrogate directly within a coupled MDO formulation to efficiently propagate uncertainty in wave conditions. In this framework, the power output of the WEC array is represented by a polynomial approximation that maps the uncertain wave environment at the deployment site to system performance, enabling the optimizer to evaluate the expected absorbed power with high fidelity. This non-intrusive formulation is particularly advantageous because the underlying WEC hydrodynamic-control model is computationally expensive and not amenable (yet) to local sensitivity or adjoint-based methods. By integrating control optimization directly into the uncertainty-aware MDO process, the approach achieves computationally efficient, risk-neutral designs that remain robust across a wide range of environmental conditions.

The remainder of this paper is organized as follows. Section~\ref{sec:probfor} introduces the simulation model and the risk-neutral optimization formulation. Section~\ref{sec:modmeth} presents the disciplinary models and methods for geometry, hydrodynamics, dynamics, control, and the regression-based surrogate used to estimate expected power. The optimization strategy and its implementation within the MDO framework are described in Section~\ref{sec:wec_array_opt}. Section~\ref{sec:wec_array_results} reports the deterministic and uncertainty-aware optimization results, followed by a discussion of their implications for WEC farm design in Section~\ref{sec:wec_array_disc}. Future research directions are outlined in Section~\ref{sec:wec_array_futurework}, and Section~\ref{sec:wec_array_concl} concludes the chapter with a summary of key findings.


\section{Problem Formulation}
\label{sec:probfor}

In an MDO framework, each discipline, such as hydrodynamics, control, and cost, is modeled as a coupled module. The overall problem includes one or more objectives and constraints, with interdependent design variables shared among disciplines. As described by \citet{agte2010mdo}, MDO is a ``quantitative design methodology that utilizes heuristic and gradient-based optimization algorithms to solve complex engineering design problems involving multiple coupled disciplines." The inter-dependencies among modules are represented through coupling variables that form both feedforward and feedback loops.  

A multidisciplinary feasible (MDF) architecture is adopted here, in which the coupling between disciplines is resolved by iterating until a feasible solution satisfying all disciplinary analyses is found ~\citep{cramer1994problem}. This strategy provides a physically consistent model to the optimizer, ensuring that each design iteration reflects an equilibrium among the interacting subsystems. 

In this study, rather than optimizing mean power alone, we seek to ensure that the WEC farm performs reliably under environmental variability due to uncertain wave conditions. The design objective is to maximize the expected power per unit volume, where volume serves as a proxy for cost since structural and material requirements dominate WEC capital cost~\citep{rm3report} and a good proxy for LCOE (levelized cost of electricity) given that the costs are uncertain in early design iterations~\citep{AZAD2025120183}. 

\subsection{Design variables and objective}

The design vector $\mathbf{x}$ includes geometric, layout, and control parameters that jointly define the hydrodynamic and PTO behavior of the array. The objective function evaluates expected electrical power normalized by total displaced volume, ensuring designs that are both energy‐efficient and materially efficient. 

\begin{figure}[h]
    \centering
    \includegraphics[width=\linewidth]{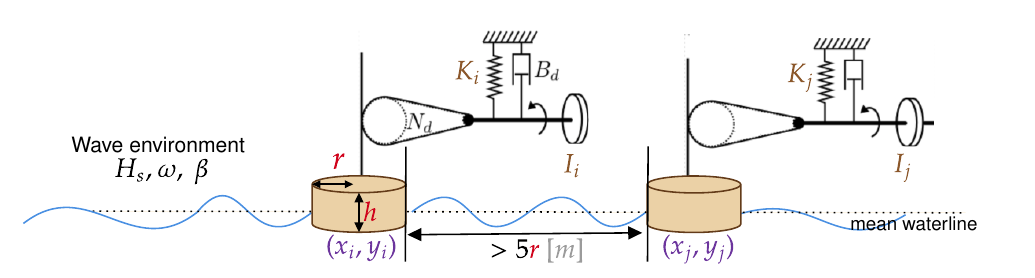}
    \caption{Schematic of the WEC geometry and control design variables. The WEC buoy radius and height are denoted by $r$ and $h$, respectively. Each WEC has individually tuned PTO parameters: drivetrain inertia $I_i$ and stiffness $K_i$. The drivetrain gear ratio, $N_d$ and linear friction $B_d$ are constant across all WECs. Adapted from \cite{Stroefer2023}.}
    \label{fig:wec_arrays}
\end{figure}

The input design vector is
\begin{equation}
    \mathbf{x} = [2r,h/r,I_{1},K_{1},x_2,y_2,I_{2},K_{2},x_3,y_3,I_{3},K_{3}, ..., x_n,y_n,I_{n},K_{n}]^{T} 
    \label{eq:x}
\end{equation}
where $r$~[m] and $h$~[m] are the radius and height of the WEC buoy, respectively, $I_i$~[ kg$\cdot$m$^2$] and $K_i$~[Nm/rad] are the PTO drivetrain inertia and stiffness for the $i$th WEC, respectively, and the subscript $n$ represents the total number of WECs. The WEC coordinates are represented by variables ($x_i,y_i$). Figure \ref{fig:wec_arrays} shows the WEC geometry and control variables for the case of a two body WEC array. 

The design variables and parameters are detailed in Tables \ref{tab:dvsccd} and \ref{tab:parameters}, respectively.

\begin{table}[h!]
\centering
\caption{Design variables, units, and bounds.}
\begin{tabular}{l c c c c}
\toprule
\textbf{Design variable} & \textbf{Units} & \textbf{Lower bound} & \textbf{Upper bound}  \\
\midrule
WEC radius ($r$) & m & 0.5 & 2.5  \\
WEC height ($L/r$) & m & 0.1$r$ & 2$r$  \\
$x$-location ($x_i$) & m & -200 & 200  \\
$y$-location ($y_i$) & m & -200 & 200  \\
PTO inertia ($I_i$) & kg$\cdot$m$^2$ & 0 & 260  \\
PTO stiffness ($K_i$) & Nm/rad & -150 & 150  \\
PTO forces ($x^{opt}_i$) & - & $-F_{\max}$ & $F_{\max}$  \\
WEC state ($x^{wec}_i$) & - & -- & --  \\
\bottomrule
\end{tabular}
\begin{flushleft}
\footnotesize
\textit{Note:} $x^{opt}_i$ are the PTO forces computed by the inner pseudo-spectral optimal control solver and are constrained to lie within $\pm F_{\max}$.  
$x^{wec}_i$ denotes the coefficients for WEC states (e.g., position) returned by the inner pseudo-spectral solver as well. All other variables are tuned by outer optimizer.
\end{flushleft}
\label{tab:dvsccd}
\end{table}

\begin{table}[h!]
\centering
\caption{Parameters.}
\begin{tabular}{l c c c}
\toprule
\textbf{Parameter} & \textbf{Units} & \textbf{Value} & \textbf{Source}\\
\midrule
1st WEC $x$-location ($x_i$) & m & 0 & -\\
1st WEC $y$-location ($y_i$) & m & 0 & -\\
Peak period ($T_p$) & s & 11.30 &  -\\
Significant wave height ($H_s$) & m & 2.0 & - \\
Wave heading ($\beta$) & rad & 0 & -\\
Max PTO force ($F_{\max}$) & N & $2.6\times10^{5}$ & \citep{penasanchez} \\
Drivetrain linear friction ($B_d$) & Nms/rad & 1.0 & \citep{Stroefer2023} \\
Drivetrain gear ratio ($N_d$) & rad/m & 12 & \citep{Stroefer2023}\\
\bottomrule
\end{tabular}
\begin{flushleft}
\footnotesize
\textit{Note:} Wave heading $\beta$ is measured from the $+x$-axis.
Note that $H_s$ and $\beta$ for deterministic case study is not the same as for uncertain case study where they were both assumed random and obtained from the NOAA buoy station ID 46042 (detailed on Section \ref{subsec:uncert})
\end{flushleft}
\label{tab:parameters}
\end{table}

The total number of design variables (including the controller parameters and inner controller action variables) can be parameterized in terms of the number of WECs (taken to be $n=6$ for this study) and the number of frequency components (taken to be $\omega = 10$ for this study) as:
\begin{equation}
\label{eq:design_choices}
N_{\mathrm{total}} = 
\underbrace{2 \, \omega \, n}_{\text{controller}} +
\underbrace{2 n}_{\text{PTO}} +
\underbrace{2 (n-1)}_{\text{WEC locations}} +
\underbrace{2}_{\text{geometry}} = 144
\end{equation}

\begin{figure}[t!]
    \centering
    \includegraphics[width=\linewidth]{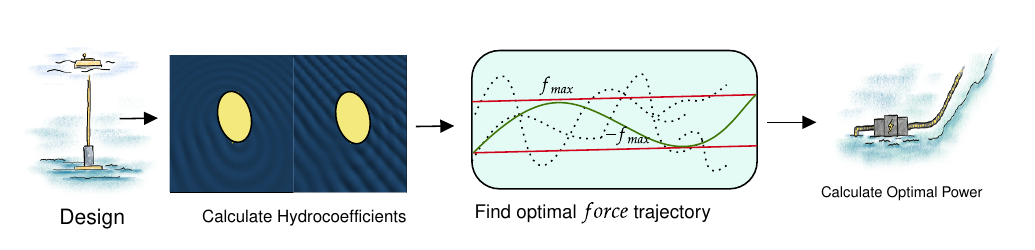}
    \caption{Conceptual illustration of the coupled hydrodynamic (wave) and control (power conversion) subsystems in the WEC array. Design variables affect both modules, and their interactions determine the total absorbed electrical power. The trajectory may not be as smoothly varying as shown in this illustration}
    \label{fig:ccdintro}
\end{figure}

These variables collectively capture the interdependence between geometry, PTO dynamics, and array layout, defining the full coupled design space in the CCD optimization of the wave-to-wire model~\citep{alves_wave--wire_2017}. In a wave-to-wire framework, the hydrodynamic wave–body interactions (``wave") are directly coupled to the drivetrain, control, and electrical power conversion models (``wire"), enabling end-to-end prediction of absorbed electrical power. A conceptual illustration of the wave-to-wire framework as used in the CCD formulation is shown in Figure~\ref{fig:ccdintro}.

\begin{figure}[t!]
    \centering
        \includegraphics[width=\linewidth]{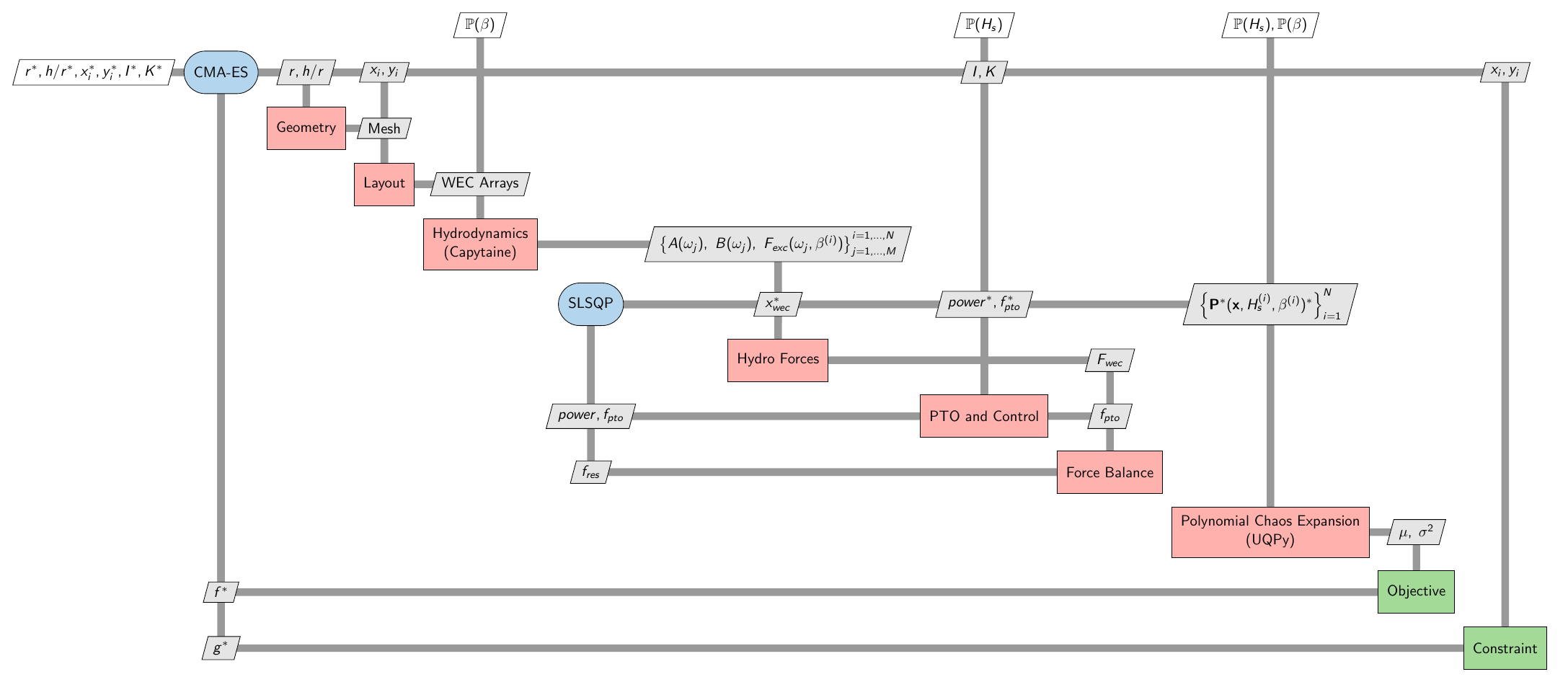}
        \caption{xDSM diagram of the uncertain multidisciplinary model of WEC farms. This architecture couples the optimizer with the multidisciplinary analysis of all the relevant modules. The diagonals are the disciplines and off-diagonals are the coupling variables. The input design variables are shown passing from the top and output of each discipline is shown on the right. The optimal value for each discipline is shown with an asterisk(*) of the left after the optimizer converges. The grey line shows the data I/O and blue line shows the sequence of analysis. }
    \label{fig:xdsm}
\end{figure}

The xDSM diagram~\citep{xdsm} is shown in Figure \ref{fig:xdsm}. It illustrates the simulation models, the disciplinary analyses involved, and the flow of interactions among them. All coupling variables are functions of the design vector, either explicitly or implicitly. The disciplines are ordered to eliminate feedback loops in the diagram.

\subsection{Treatment of uncertainty}
\label{subsec:uncert}

The performance of WEC arrays is strongly influenced by variability in ocean wave conditions. This uncertainty is classified as aleatoric, arising from the inherent stochastic nature of the ocean environment. Epistemic uncertainty due to model limitations is not considered in this study but could be mitigated in future work by incorporating experimental data~\citep{UQ_review}.

\citet{AZAD2025120183} optimized the expected (weighted average) absorbed power of a WEC farm using a probability distribution of sea states derived from historical wave data~\citep{ruiz2017hydrodynamic}. In contrast to that study, which employed simplified fixed control strategies for discrete sea states, our formulation incorporates constrained optimal control with realistic WEC dynamics. 

To characterize the environmental uncertainty, we fit marginal probability
distributions to significant wave height ($H_s$), mean wave direction
($\beta$), and peak wave period ($T_p$) using observations from the NOAA
Monterey Bay buoy (station ID 46042) for the year 2023. The significant wave height, $H_s$~[m], is modeled using a Rayleigh distribution characterized by scale parameter $\sigma_{H_s}$, while the wave direction, $\beta$~[deg], is modeled as a truncated normal distribution on $0^\circ$ and $360^\circ$. The peak period, $T_p$~[s], is modeled using a two-parameter Weibull distribution with shape
parameter $k_{T_p}$ and scale parameter $\lambda_{T_p}$. To capture the
statistical dependence among these variables, a Nataf transformation is used to generate correlated random samples that preserve both the fitted marginals and their empirical correlation structure.

\begin{figure}[h]
    \centering

    \begin{subfigure}{0.48\textwidth}
        \centering
        \includegraphics[width=\linewidth]{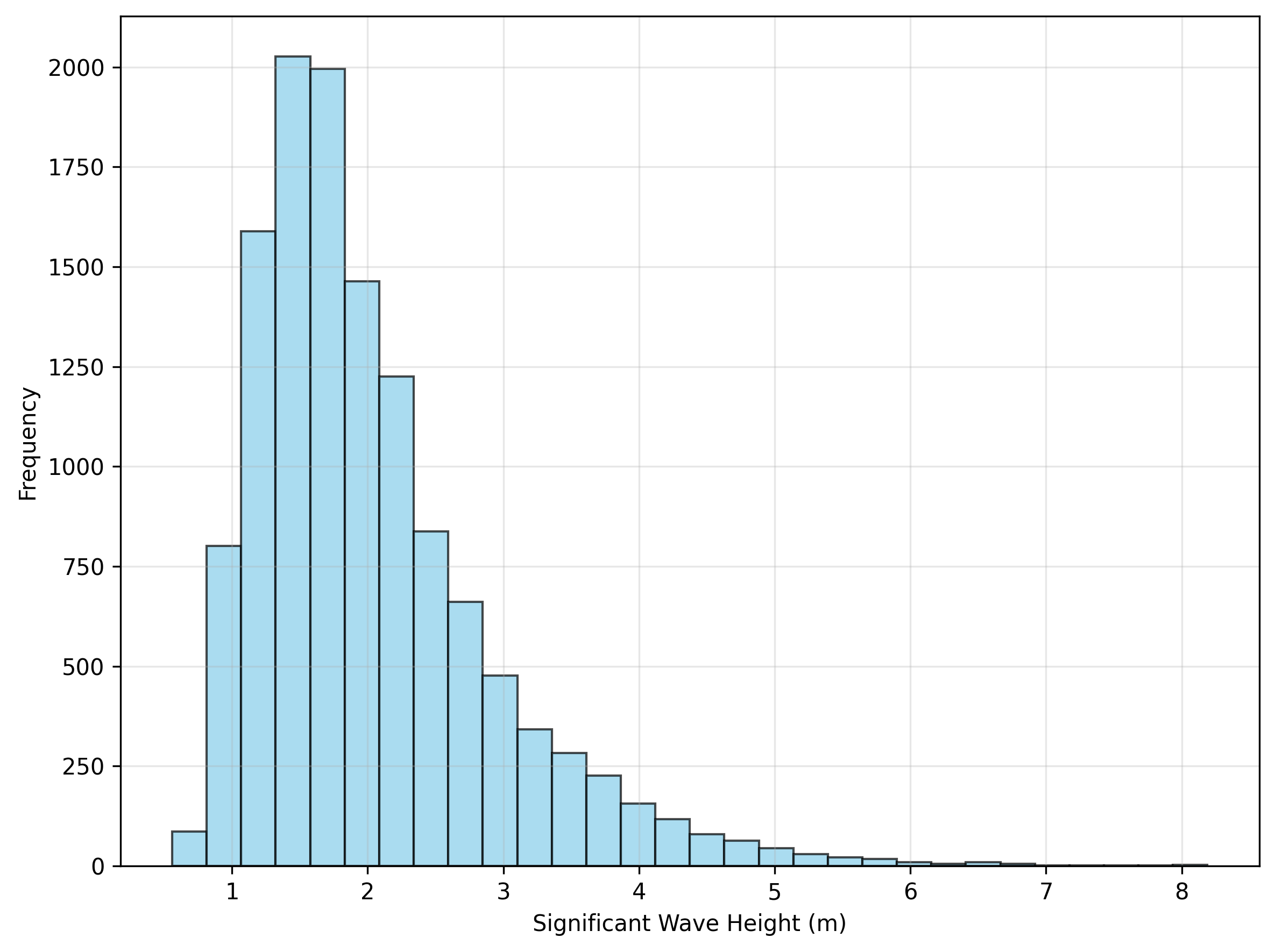}
        \caption{Significant wave height ($H_s$).}
    \end{subfigure}
    \hfill
    \begin{subfigure}{0.48\textwidth}
        \centering
        \includegraphics[width=\linewidth]{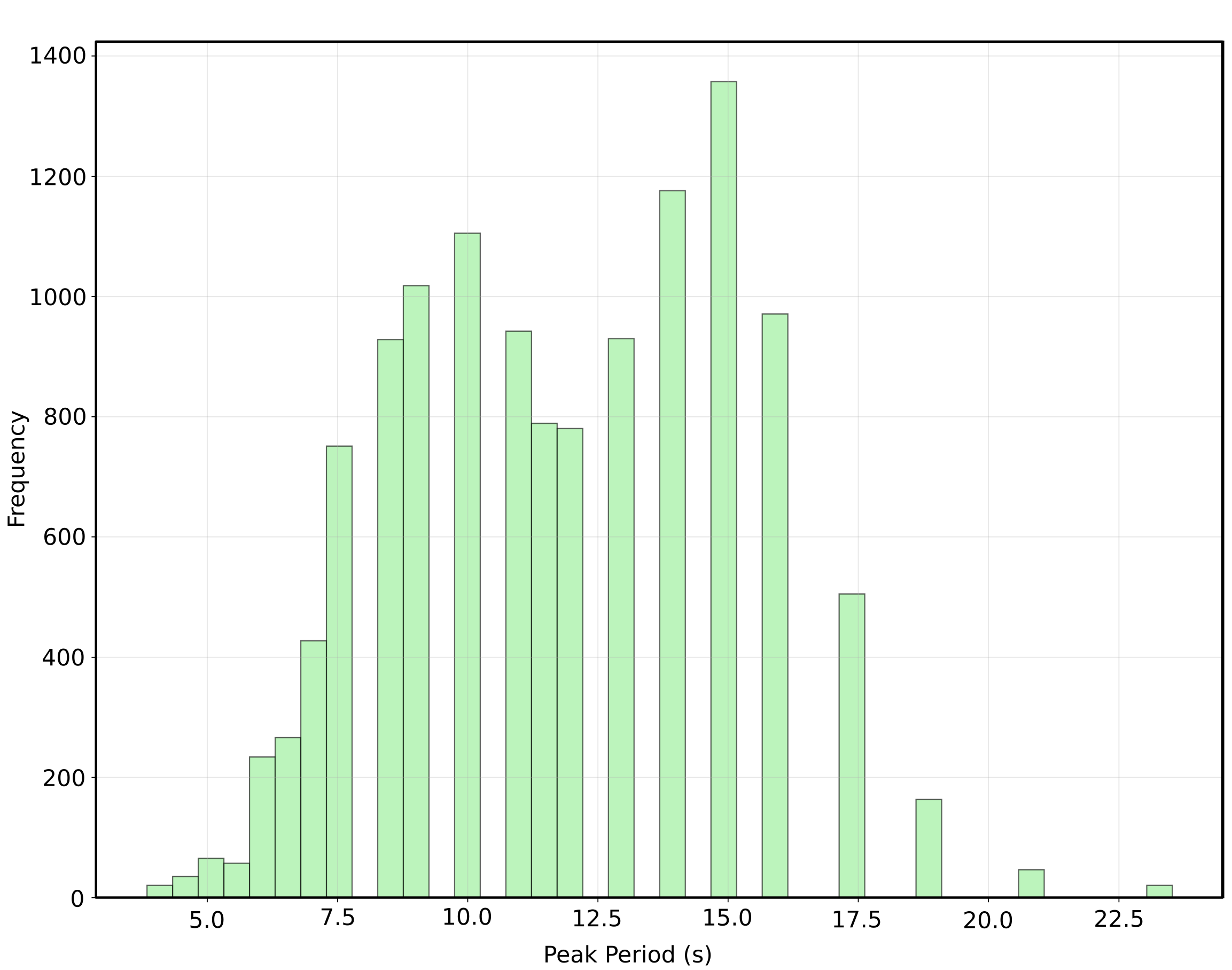}
        \caption{Peak period ($T_p$).}
    \end{subfigure}

    \vspace{0.5em}

    \begin{subfigure}{0.48\textwidth}
        \centering
        \includegraphics[width=\linewidth]{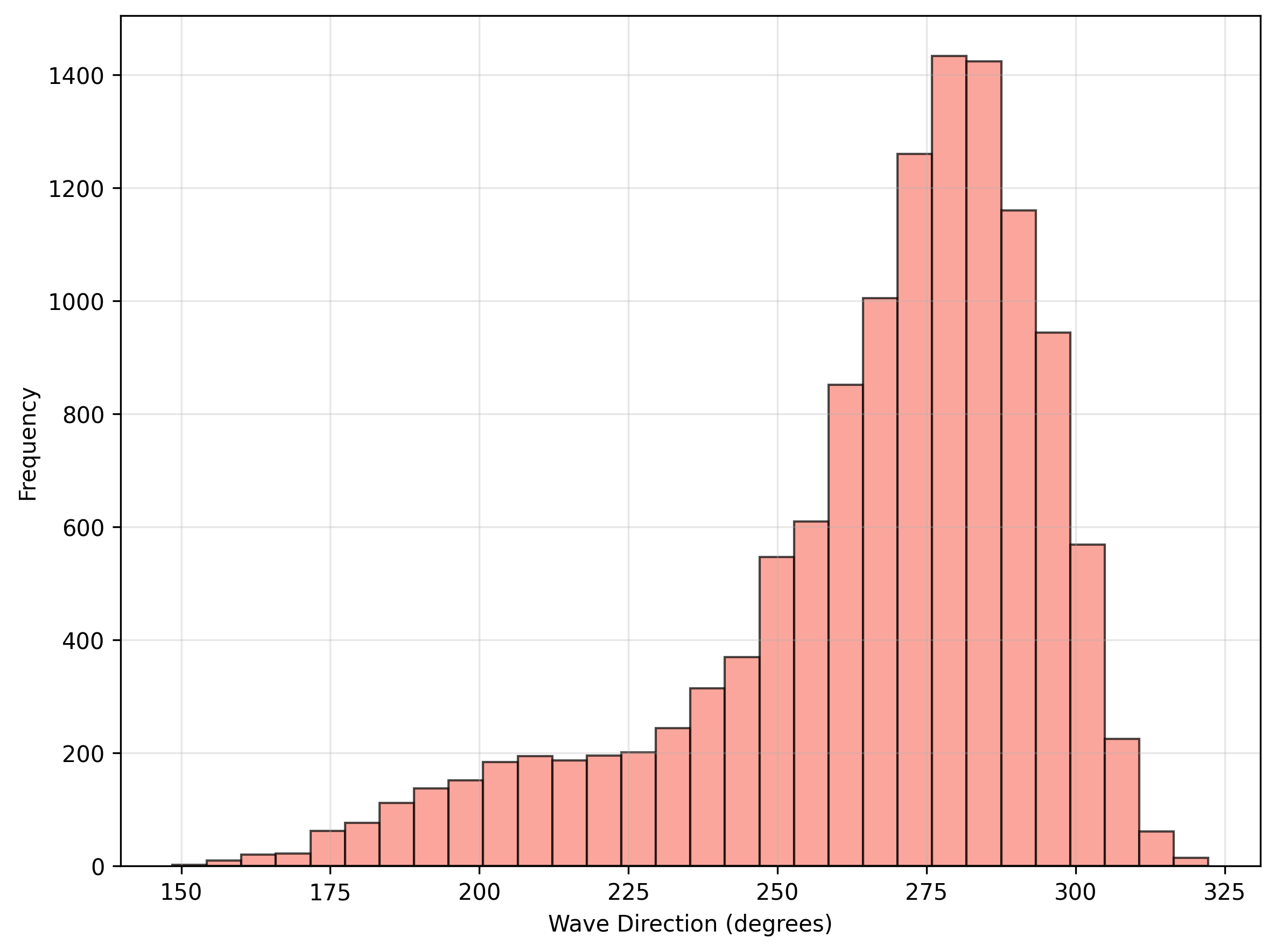}
        \caption{Wave direction ($\beta$).}
    \end{subfigure}

    \caption{Distributions of (a) significant wave height, (b) peak period, and (c) mean wave direction for Monterey Bay, California (NOAA buoy ID 46042, year 2023).}
    \label{fig:distributionofspectrum}
\end{figure}

The expected absorbed power is then expressed as $\mathbb{E}[P(\mathbf{x}; H_s, T_p, \beta)]$, where the expectation is taken over these probability distributions fitted to the empirical data shown in Figure~\ref{fig:distributionofspectrum}.

\subsection{Formal mathematical problem statement}

The MDO problem seeks a robust configuration of the WEC farm that maximizes expected electrical power production per structural volume under uncertain sea states. The design variables $\mathbf{x}$ include geometry, layout, and control parameters, and the uncertain environmental inputs $(H_s, T_p \beta)$ are modeled probabilistically as described in Section~\ref{subsec:uncert}.

The formal optimization problem is expressed as:
\begin{align}
    \text{maximize}_{\mathbf{x}} \quad 
    & J(\mathbf{x}) = \mathbb{E}_{H_s,T_p,\beta}\Bigg[\frac{P(\mathbf{x}; H_s, T_p, \beta)}{V(\mathbf{x})}\Bigg]  \label{eq:objective}\\[4pt]
    \text{subject to} \quad 
    & \sqrt{(x_i - x_j)^2 + (y_i - y_j)^2} - 5r \ge 0, \quad \forall i \ne j, \label{eq:spacing}\\[4pt]
    & R_i(\mathbf{x}, \mathbf{p}; H_s, \beta) = 0, \label{eq:residual}\\[4pt]
    & \mathbf{x}_{\min} \le \mathbf{x} \le \mathbf{x}_{\max}, \label{eq:bounds}\\[4pt]
      & |f^i_{pto}| \leq F_{max}\\[4pt]
    & H_s \sim \text{Rayleigh}(\sigma_{H_s}=0.889~\mathrm{m}), \label{eq:hs_dist}\\
     & T_p \sim \text{Weibull}(c = 3.918,\; \lambda = 12.888\,\text{s}), \label{eq:tp_dist}\\
    & \beta \sim \text{TruncNormal}(\mu_\beta=266.46^\circ,\, \sigma_\beta=29.42^\circ;\, 0^\circ,\, 360^\circ). \label{eq:beta_dist}
\end{align}

Here, $\mathbb{E}_{H_s,T_p,\beta}\big[P(\mathbf{x}; H_s, T_p, \beta)\big]$~[W/m$^3$] denotes the risk-neutral expected power output, computed over the joint probability distribution of the significant wave height $H_s$, wave period ($T_p$) and mean wave direction $\beta$. The constraint in Equation~\eqref{eq:spacing} enforces a minimum spacing of $5r$ between devices, ensuring physical feasibility and preventing hydrodynamic overlap while allowing meaningful array interactions~\citep{BABARIT201368}.

Equation~\eqref{eq:residual} represents the implicit system residuals coupling geometry, hydrodynamic, dynamics and controls as outlined in Section~\ref{sec:mdo}. Bounds in Equation~\eqref{eq:bounds} define the feasible design region for all variables listed in Table~\ref{tab:dvsccd}. The resulting formulation captures the coupling between design and control variables explicitly and implicitly through the objective function $P(\mathbf{x}; H_s, T_p, \beta)$.

In this formulation, the optimization framework integrates uncertainty quantification directly into the objective evaluation. For each design vector $\mathbf{x}$, the power response $P(\mathbf{x}; H_s, T_p, \beta)$ is approximated via a PCE surrogate model. This approach enables efficient computation of the expected performance $\mathbb{E}[P]$ while maintaining fidelity to the fully coupled hydrodynamic-control dynamics. 

By coupling the geometric and control design~\citep{AZAD2025120183} and computing expectation across the vary parameters, the problem formulation ensures that optimized designs exhibit both high performance and statistical robustness to environmental variability, in contrast to traditional deterministic sequential optimization methods

\section{Modeling and Methods}
\label{sec:modmeth}
In this section, we discuss the various disciplines and the governing equations in each including geometry, layout, hydrodynamics, dynamics and control, and power.
\label{sec:mdo}

\subsection{Discipline Models}

\subsubsection{Geometry}

The geometry module generates identical cylindrical WECs distributed across the array, as shown in Figure~\ref{fig:wec_arrays}. Each buoy is modeled as a vertical cylinder designed to oscillate in heave about its center of mass. The geometric parameters determine the hydrodynamic properties of each device, while the spatial arrangement governs the wave interaction (constructive and destructive) and energy sharing among units. Because both the shape of individual converters and their relative positioning strongly influence farm-level performance, they are taken as a design variables in the optimization framework.

Beyond individual device sizing, the spatial layout of the array has an effect on collective performance. Hydrodynamic interactions among devices modify the wave fields, producing constructive or destructive interference patterns that directly affects the total absorbed power.

In this study, a heaving point absorber WEC is considered, with cost and performance directly influenced by its geometry and size. The geometry is parameterized by two key design variables: the buoy radius $r$~[m] and draft $h$~[m]. Together, these  determine displaced volume, hydrostatic stiffness, and total structural mass. In conventional WEC design, these parameters are usually optimized for a given wave environment to maximize energy absorption. Here, they are embedded within a coupled MDO framework to capture their broader interactions with array layout and control performance. These geometric properties affect not only the energy absorption potential but also material and fabrication costs, making them critical to the overall optimization.

\subsection{Layout}
The layout model defines the physical configuration of the WEC farm. The layout is defined by the Cartesian coordinates $(x_i, y_i)$ of each device, which are treated as continuous design variables in the optimization. Unlike approaches that fix the overall array topology, this formulation allows the optimizer to freely explore arbitrary spatial configurations rather than being restricted to predefined patterns (although with increased computational cost). The initial layout is varied within the design domain while enforcing a minimum spacing constraint of $5r$ between neighboring buoys to prevent overlap and maintain hydrodynamic validity. This flexible parameterization captures the key geometric and spatial degrees of freedom required for integrated geometry-layout optimization within the MDO framework.

\citet{YANG2022112668} surveyed different layout topologies, including regular and staggered grids, highlighting the computational benefits of optimizing layout type and inter-device spacing to reduce the dimensionality of the problem. \citet{gotemanReview} analyzed over 1000 interacting point absorber WECs using an approximate analytic method for single body diffraction and found that while mean absorbed power remained comparable across configurations (wedge, rectangular, circular, random), the variance in absorbed power differed substantially, suggesting that layout primarily influences the robustness of array performance rather than its mean efficiency. Using a higher-fidelity boundary element method (BEM) hydrodynamics solver, \citet{BORGARINO201279} further showed that power absorption is relatively insensitive to device spacing when PTO damping is optimally tuned, indicating that separation distance alone has limited influence once control effects are considered.

\subsubsection{Hydrodynamics}

The hydrodynamic model characterizes the interaction between the incident ocean waves and the oscillating WEC bodies and their response. Building on the geometric definitions established in the previous section, this model evaluates how device shape, size, and spatial configuration influence wave radiation, diffraction, and excitation forces within the array. The resulting hydrodynamic coefficients (added mass, damping, and excitation force) form inputs to the subsequent dynamics and control models. Accurate hydrodynamic modeling is therefore critical for estimating system performance.

In this study, wave-structure interactions are described using linear potential flow theory, which assumes the fluid is inviscid, incompressible, and irrotational. The velocity field $\vec{v}$ is expressed as the gradient of a complex velocity potential $\phi$:

\begin{equation}
    \vec{v} = \nabla \phi. 
\end{equation}
Because the flow is irrotational, the potential $\phi$ satisfies the Laplace equation: 
\begin{equation}
\nabla^2 \phi = 0.
\end{equation}

The time-harmonic complex velocity potential is written as 
\begin{equation}
\Phi = \mathbb{Re}(\phi e^{-j\omega t}),
\end{equation}
where $\omega$~[rad/s] is the wave frequency. The boundary value problem (BVP) is closed by applying the linearized free-surface conditions, no-flux conditions on the seabed and body surface, and an outgoing-wave radiation condition at infinity~\citep{Falnes_2002}.

The hydrodynamics module also computes the hydrostatic stiffness coefficient and WEC buoy mass. The hydrostatic stiffness in heave is
\begin{equation}
C_{33} = \rho g A_{wp}
\end{equation}
where $A_{wp}$~[m$^2$] is the waterplane area, and the WEC mass is 
\begin{equation}
    M_{bb} = \rho g V_b,
\end{equation}
where $\rho$~[kg/m$^3$] is the seawater density, $g$~[m/s$^2$] is the gravitational acceleration, and $V_b$~[m$^3$] is the submerged volume. Together, these quantities define the inertial and restoring properties of the floating body.

Most hydrodynamic solvers for wave-body interaction problems (incident, diffraction and radiation) based on reformulating of the governing Laplace equation and its boundary conditions as boundary integral equations (BIEs), which are then solved using the boundary element method (BEM). By applying the Green's identity, the volume problem is converted into a surface-only formulation, requiring discretization of the wetted body surface rather than the entire fluid domain~\citep{bem_wave}. In practice, the wetted surface is meshed into boundary panels, and the resulting BIE is discretized into a linear system of equations relating the unknown velocity potential at each panel to the prescribed boundary conditions. This surface-based formulation is particularly well suited for unbounded ocean domains, complex or irregular geometries, and multi-body WEC arrays, as it naturally enforces the radiation condition at infinity and captures hydrodynamic interactions between multiple bodies efficiently. The reduced discretization effort and computational cost make BEM advantageous in design optimization studies that require repeated evaluations across varying WEC sizes, shapes, and array configurations.

In this study, hydrodynamic coefficients for each array configuration are computed using the open-source frequency-domain BEM solver Capytaine~\citep{ancellin_capytaine_2019, babarit_theoretical_2015}. The code implements the linear potential flow formulation described above and solves the boundary integral system for the discretized buoy surfaces. Its modular, Python-based architecture facilitates efficient parametric analysis and integration with the optimization framework.

When multiple WECs are placed near to each other, the radiated and diffracted wave fields interact, producing constructive or destructive interference that alters the energy absorbed by each device. As shown by \citet{Budal1977}, the total absorbed energy in an array does not scale linearly with the number of bodies and may differ for various array configurations due to these interactions. The hydrodynamic interactions depend on the device geometry, spacing, and incident wave direction, making it essential to analyze hydrodynamics and layout together in a coupled framework.  These interactions are important to consider in early design studies.  

\begin{figure}[t!]
    \centering
    \begin{subfigure}{0.45\textwidth}
        \centering
        \includegraphics[width=\textwidth]{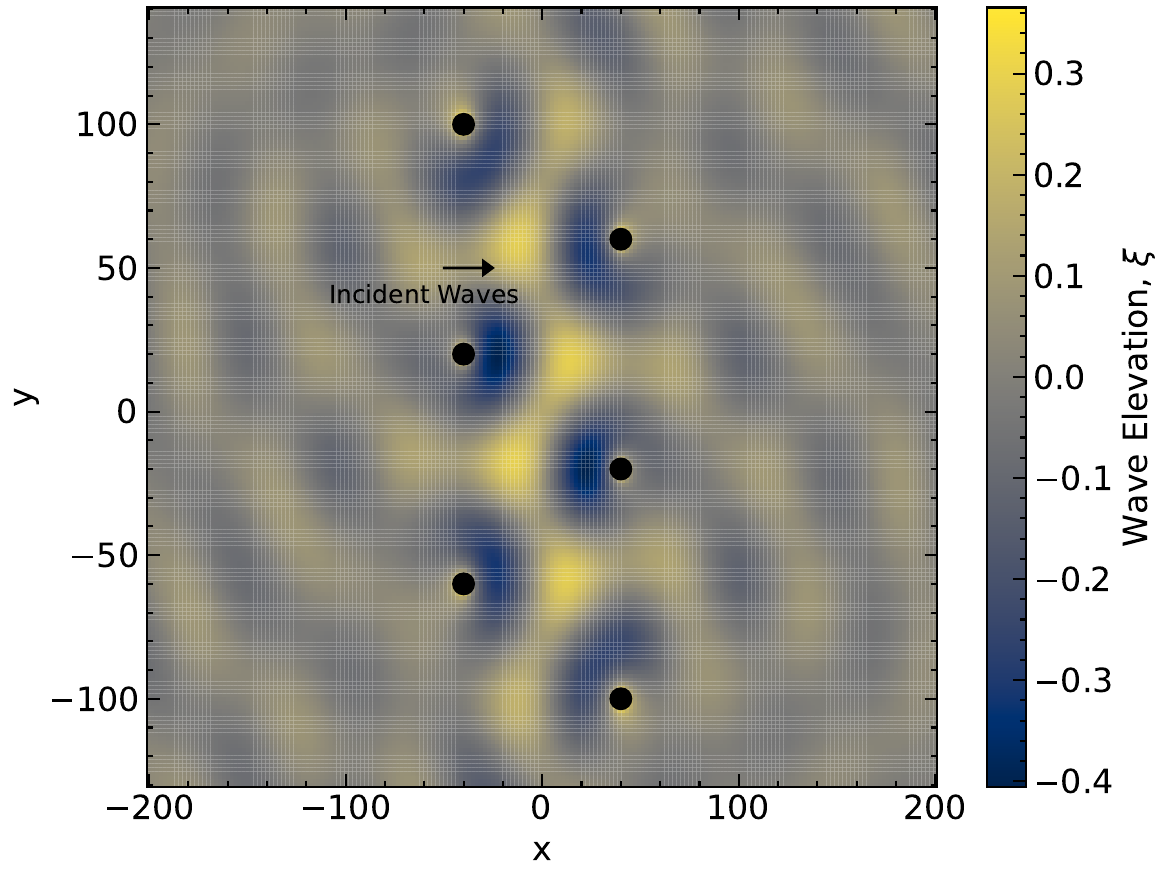}
        \caption{The radiated wave elevation.}
        \label{fig:rad}
    \end{subfigure}
    \hfill
    \begin{subfigure}{0.45\textwidth}
        \centering
        \includegraphics[width=\textwidth]{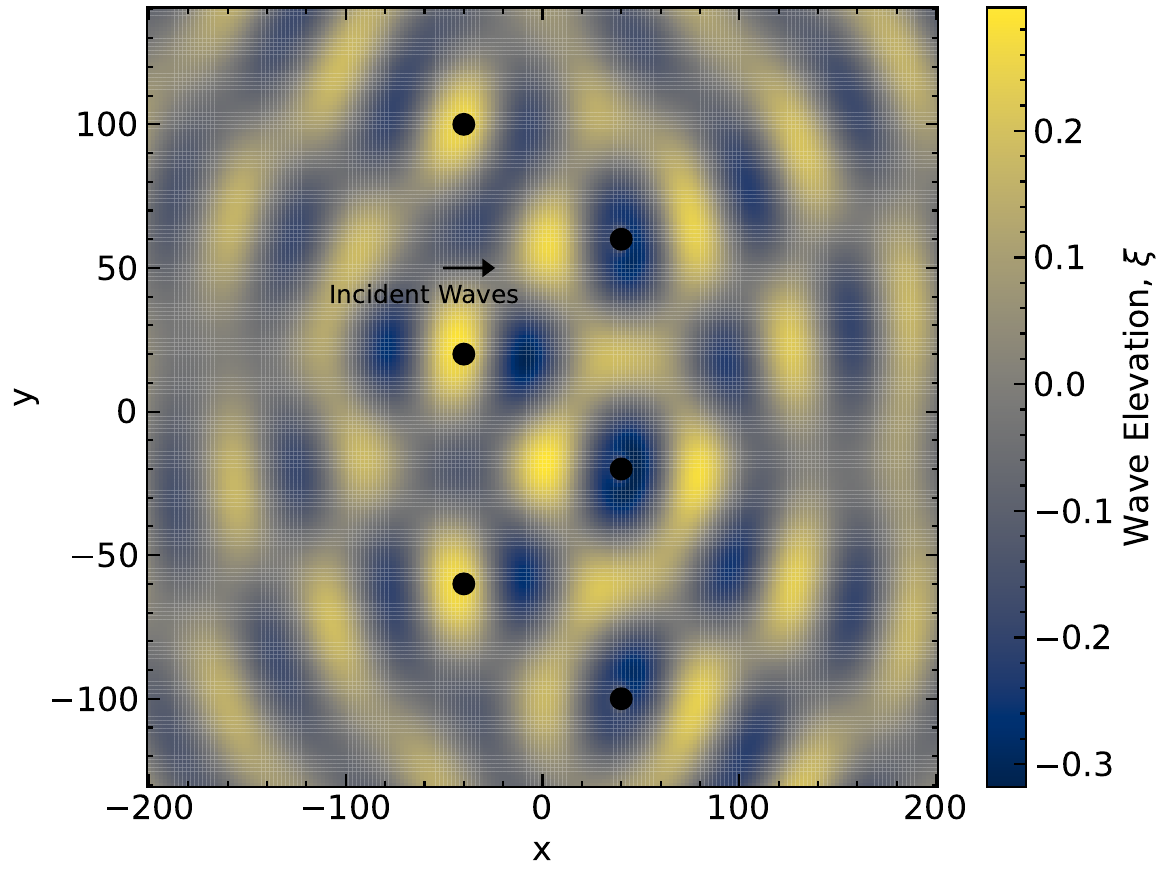}
        \caption{The diffracted wave elevation.}
        \label{fig:diff}
    \end{subfigure}
    \hfill
    \begin{subfigure}{0.45\textwidth}
        \centering
        \includegraphics[width=\textwidth]{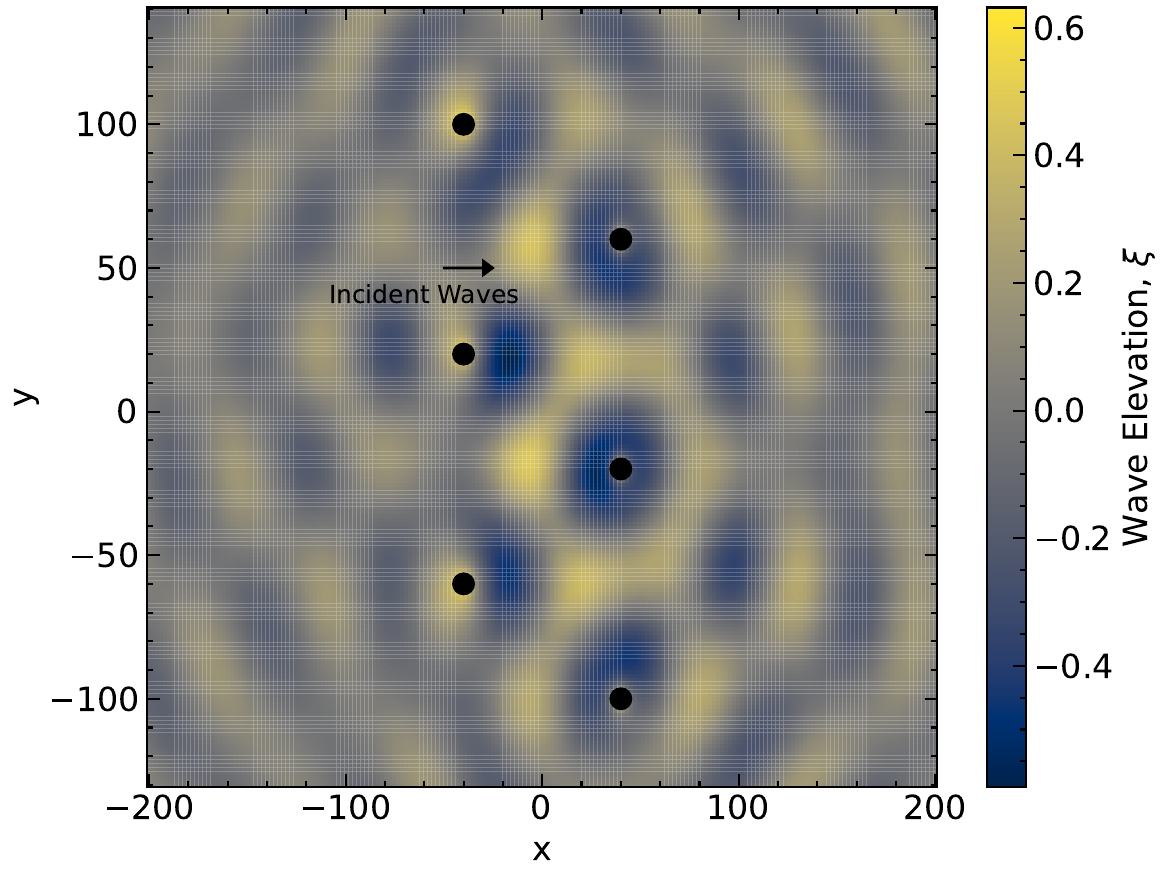}
        \caption{Combined (a+b) wave elevation.}
        \label{fig:rad_dif}
    \end{subfigure}
    \hfill
    \begin{subfigure}{0.45\textwidth}
        \centering
        \includegraphics[width=\textwidth]{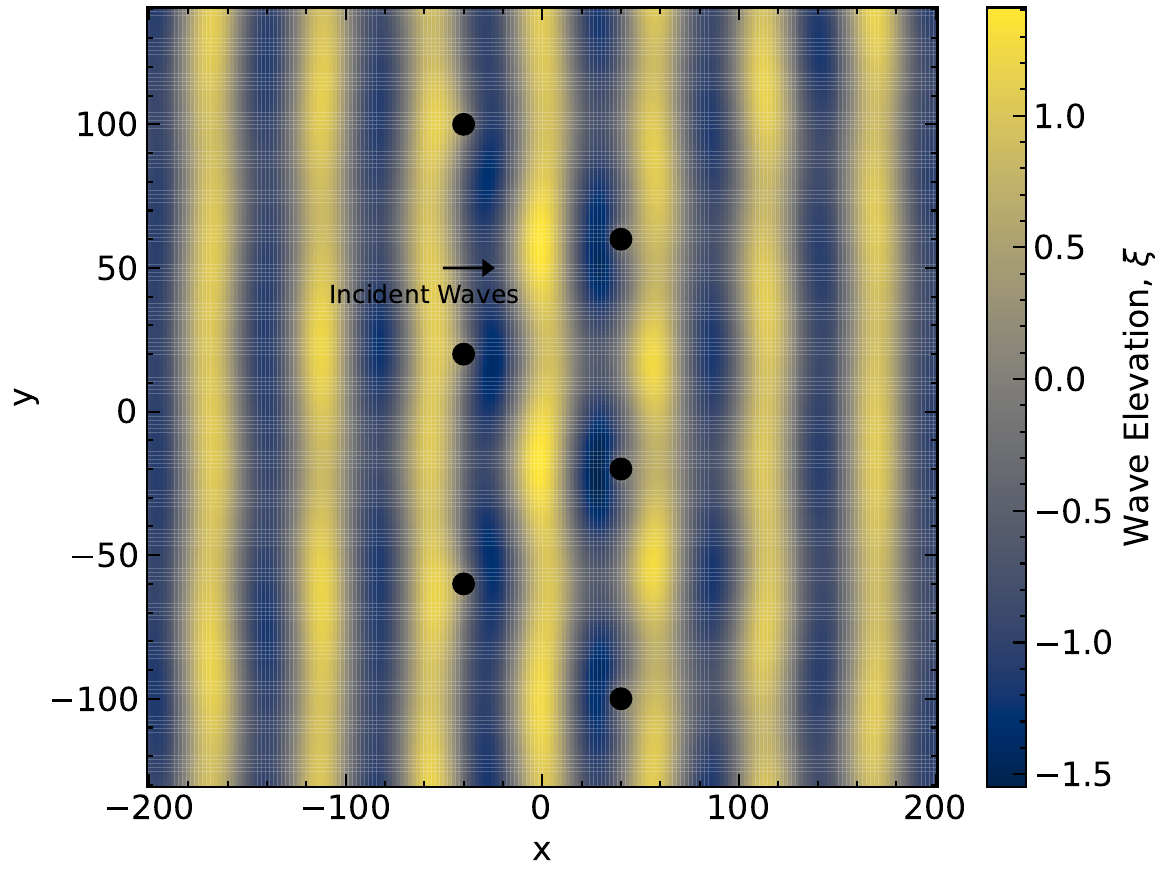}
        \caption{The total wave elevation.}
        \label{fig:total}
    \end{subfigure}
    \caption{The elevation of the radiated, diffracted, and total wave fields for an arbitrary 6-body WEC array showing the wave interference patterns with regular waves of unit amplitude and frequency of 1.04 [rad/s]. }
    \label{fig:wave elevation}
\end{figure}

Various approaches exist for modeling array interactions, including analytical~\citep{simon1982multiple}, semi-analytical~\citep{zhong2016wave, mcnatt2015novel, singh2013hydrodynamic}, and predictive methods~\citep{ZHU2022112072}. In this study, full BEM resolution is employed to explicitly capture all pairwise interactions among panels across devices. The diffraction and radiation problems are solved in the frequency domain to compute the corresponding velocity potentials $\phi_{\text{diffraction}}$ and $\phi_{\text{radiation}}$, yielding the total potential field
\begin{equation}
    \phi_{\text{total}} = \phi_{\text{incident}} + \phi_{\text{diffraction}} + \phi_{\text{radiation}}.
\end{equation}
The wave elevations of the radiated, diffracted, and total wave fields for an example array are shown in Figure~\ref{fig:wave elevation}. The diffraction analysis is repeated for each wave heading $\beta$, further motivating the integration of hydrodynamic analysis within the overall MDO loop.

The radiation potential $\phi_{\text{radiation}}$ is used to evaluate the frequency-dependent hydrodynamic coefficients: added mass $\mathbf{A}_{6\times6}$ and radiation damping $\mathbf{B}_{6\times6}$, given by
\begin{equation}
    \mathbf{A}_{ij} - \frac{j}{\omega}\mathbf{B}_{ij} = - \rho \iint_{\Gamma} n_i \phi_{\text{radiation}} \, ds,
\end{equation}
where $\omega$~[rad/s] is the wave frequency, $\Gamma$ is the immersed body surface and $\rho$~[kg/m$^3$] is the fluid density, and $n_i$ is the normal vector. The wave-excitation force $\vec{F}_{1\times6}$ is determined from the pressure field associated with the diffracted potential:
\begin{equation} \label{eq:excitation_force}
    \vec{F} = -\omega \rho \iint_{\Gamma} n_i \phi_{\text{diffraction}(\beta)} \, ds,
\end{equation}
where $\phi_\text{diffraction}(\beta)$ is the potential for wave incident at angle $\beta$. Both $\mathbf{A}$ and $\mathbf{B}$ depend on the wave frequency $\omega$, with $\mathbf{A}$ proportional to acceleration and $\mathbf{B}$ proportional to velocity.

Full BEM resolution provides the highest accuracy for design optimization and sensitivity studies but incurs significant computational cost~\citep{TEIXEIRADUARTE2022112513}. Direct numerical evaluation scales as $O(N_p^2)$ for Green’s function computation and influence matrix assembly, and $O(N_p^3)$ for solving the resulting linear system, where $N_p$ is the total number of surface panels across all WECs in the array. Computational performance further depends on the choice of linear solver, the free-surface Green’s function formulation, and the influence matrix representation.

\begin{figure}[H]
    \centering
\includegraphics[width = \linewidth]{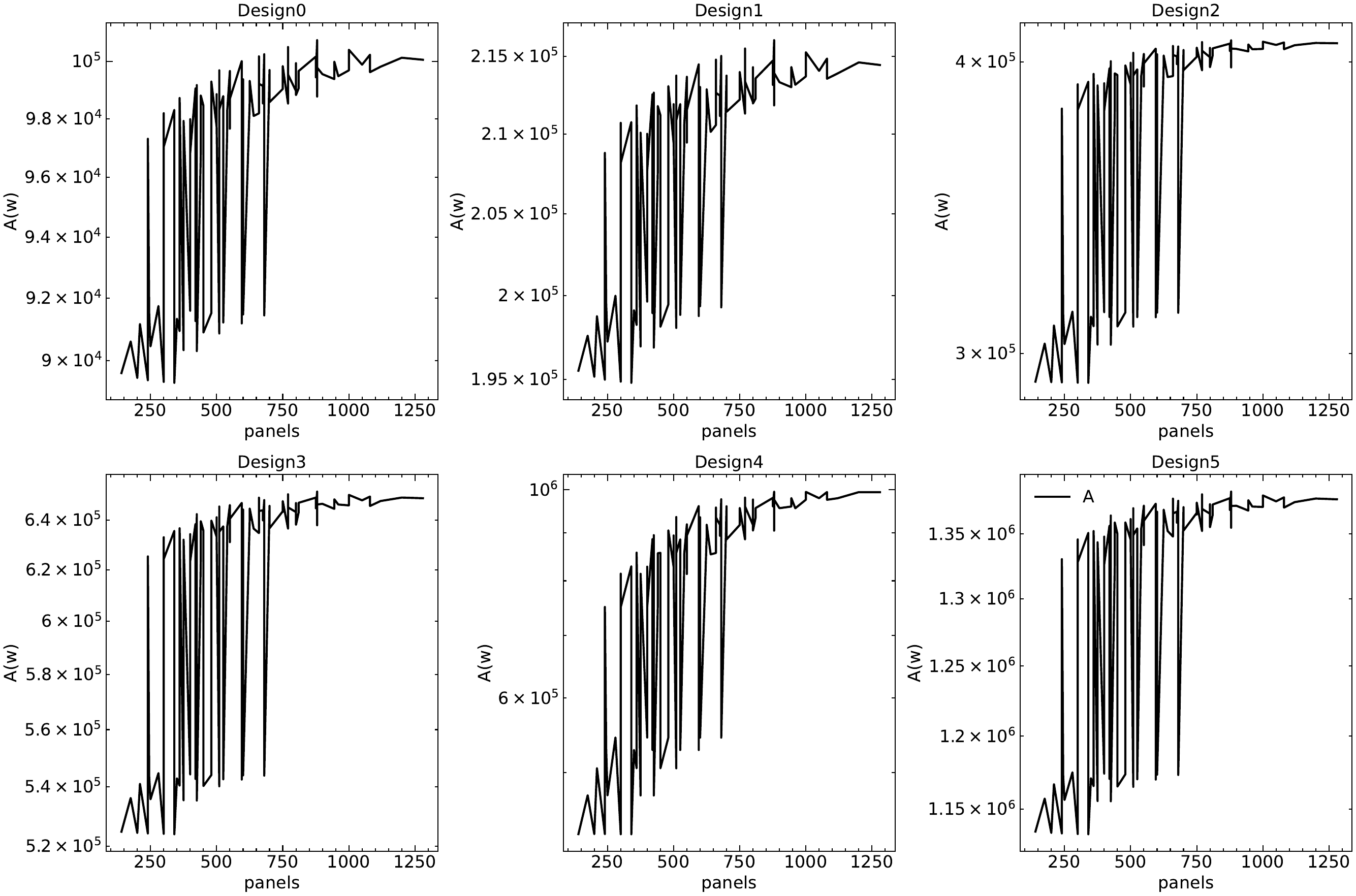}
    \caption{Mesh convergence for sample designs. The number of panel is shown on the x-axis and the $\mathbf{A}(\omega)$ is shown on the y-axis. All of these design converge for at least 750-1000 panels.}
    \label{fig:mesh_convergence}
\end{figure}

Several acceleration strategies have been proposed in prior work to reduce the computational cost of hydrodynamic simulations, including surrogate modeling, hybrid optimizers, and modified genetic algorithm operators~\citep{hybridOptimizer, GAOperator, BAILEY2018721, newman1985algorithms}. To improve efficiency within this study, a mesh convergence analysis was conducted to minimize unnecessary refinement (Figure~\ref{fig:mesh_convergence}). A representative set of designs was selected from the design space using Latin hypercube sampling ($N_S = 6$), and the resulting hydrodynamic coefficients were compared across successive mesh refinements to ensure accuracy. The final resolutions adopted for analysis are summarized in Table~\ref{tab:resolution}, where $(n_r, n_\theta, n_x)$ denote the number of panels along the radial direction, around the circumference, and along the vertical axis, respectively.

\begin{table}[h]
\centering
\begin{tabular}{|c|c|c|}
\hline
 Radius ($r$) & Resolution $(n_r, n_\theta, n_x)$ & Total Panels $N_{\text{total}}$ \\
\hline
 $> 5$        & $(7, 35, 25)$ & $ 6125$ \\
 $4 < r \leq 5$  & $(5, 25, 20)$ & $ 2500$ \\
 $3 < r \leq 4$  & $(3, 15, 15)$ & $ 675$ \\
 $\leq 3$        & $(2, 10, 10)$ & $200$ \\
\hline
\end{tabular}
\caption{Mesh resolution based on cylinder radius. $n_r$, $n_\theta$, and $n_x$ represent the number of panels along a radius at the end, the number of panels around a slice, and the number of slices, respectively.}
\label{tab:resolution}
\end{table}

\begin{figure}[H]
    \centering
     \includegraphics[width = 1.1\linewidth]{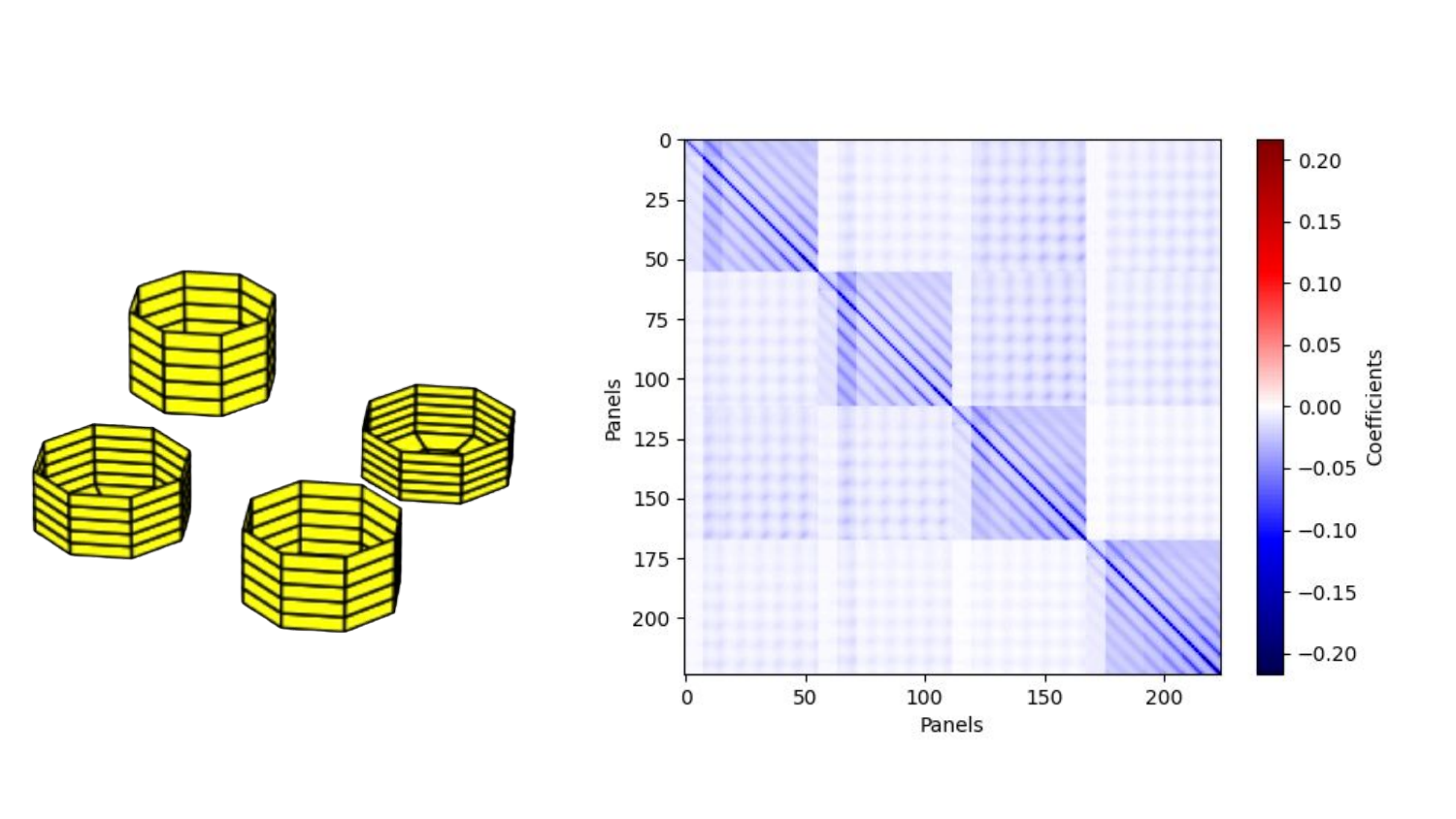} 
    \caption{Block structure in influence matrices computed via full BEM resolution, illustrating potential for low-rank compression.}
    \label{fig:Hmatrices}
\end{figure}

Further reductions in computational cost in this study are achieved through low-rank matrix compression techniques integrated into modern BEM solvers. Each block in the hydrodynamic influence matrix represents the interaction between groups of surface panels, as illustrated in Figure~\ref{fig:Hmatrices}. For distant panel groups, corresponding to off-diagonal blocks, not all matrix entries need to be computed exactly. Instead, these interactions can be efficiently approximated and compressed, substantially reducing memory requirements and accelerating the solver. \citet{Hmatrices_mathieu} demonstrated that adaptive cross approximation (ACA) can construct such low-rank representations while maintaining acceptable accuracy in BEM for marine hydrodynamics. In this study, an ACA tolerance of $10^{-1}$ and a distance threshold of seven times the buoy radius were applied. To the best of the authors’ knowledge, low-rank matrix approximation techniques such as ACA have not previously been applied within WEC array layout optimization. Given the presence of distinct low-rank block structures across different configurations, this study utilizes this as an effective acceleration strategy for large-scale MDO analyses of interacting WEC farms.

\subsubsection{Dynamics and Control}

The dynamics and control module governs the time evolution of each WEC's motion and its interaction with the PTO system. The model is formulated using linear wave-body dynamics and an impedance-based PTO representation, enabling co-design of mechanical and electrical subsystems within the coupled MDO framework.

Following \citet{falnes2020}, the equation of motion for a heaving WEC is expressed as
\begin{equation}
    \label{eq:wec}
    m\ddot{\xi} = F_e - F_r - F_b - F_v - F_f - F_u ,
\end{equation}

\noindent where $m$~[kg or kg·m$^2$] is the inertia of the WEC, and $\xi$~[m or rad] denotes its linear or angular motion. The term $F_e$~[N or Nm] represents the excitation force or torque induced by incident waves, while $F_r$~[N or Nm] is the radiation force or torque arising from the WEC’s oscillations, incorporating both added mass and radiation damping effects. The hydrostatic restoring contribution, $F_b$~[N or Nm], captures buoyancy-induced forces or torques. The drag force $F_v$~[N or Nm], resulting from nonlinear viscous effects, and the frictional term $F_f$~[N or Nm] are both neglected in this study in accordance with linear potential flow assumptions. Finally, $F_u$~[N or Nm] represents the PTO force or torque, corresponding to mechanical energy extraction through the device’s conversion system and is a function of PTO design and control actuation.

Given the oscillatory nature of wave excitation, the dynamics are conveniently expressed in the frequency domain, where each variable is represented as
\begin{equation}
    \label{eq:fd_to_td}
    x = \mathbb{X} e^{-j\omega t},
\end{equation}
with $x$ as the time-domain quantity and $\mathbb{X}$ its complex frequency-domain representation. This formulation applies to $\xi$ and all force components, facilitating frequency-by-frequency analysis of wave-structure interactions.

The PTO subsystem converts hydrodynamic motion into electrical power, acting as the link between mechanical and electrical components, and is modeled using a two-port impedance formulation~\citep{Stroefer2023}:
\begin{equation}
    \begin{bmatrix} \mathbb{F}_{u}(\omega) \\ \mathbb{V}(\omega) \end{bmatrix} 
    = 
    \begin{bmatrix} Z_{F\Xi}(\omega) & Z_{FI}(\omega) \\ Z_{V\Xi}(\omega) & Z_{VI}(\omega) \end{bmatrix}
    \begin{bmatrix} j\omega \Xi(\omega) \\ \mathbb{I}(\omega) \end{bmatrix},
    \label{eq:pto}
\end{equation}
where $\mathbb{F}_{u}$ is the complex frequency-domain representation of the PTO force $f_u$, $\mathbb{V}$ is the electrical voltage, and $j\omega \Xi$ represents the derivative of the motion in the frequency domain (i.e., the body velocity), $\mathbb{I}$ denotes the electrical current, and $Z_{F\Xi}$, $Z_{FI}$, $Z_{V\Xi}$, and $Z_{VI}$ correspond to the individual impedance components of the PTO system. These impedances are frequency-dependent functions of the PTO design parameters, most notably the stiffness and damping terms ($K_i$, $I_i$), and follow the formulation of \citet{Stroefer2023}. This representation provides a compact description of the coupled electro-mechanical dynamics, enabling efficient integration with the hydrodynamic model.

The coupled dynamics and control problem is solved using the pseudo-spectral optimal control solver WecOptTool~\cite{Stroefer2023,bacelli_numerical_2015}. The solver performs gradient-based optimization of the control trajectory while enforcing the equation of motion given in Equation~\eqref{eq:wec} as an equality constraint. This formulation treats both the control trajectory and the WEC motion as design variables within the nested optimization problem. Although it may seem counterintuitive to optimize the motion directly, doing so allows the dynamic equilibrium constraint, defined by the equation of motion, to be explicitly enforced throughout the solution process. Although the force balance is evaluated in the time domain, the motion variables are represented as Fourier coefficients.

\begin{figure}[h]
    \centering
    \includegraphics[width=0.5\linewidth]{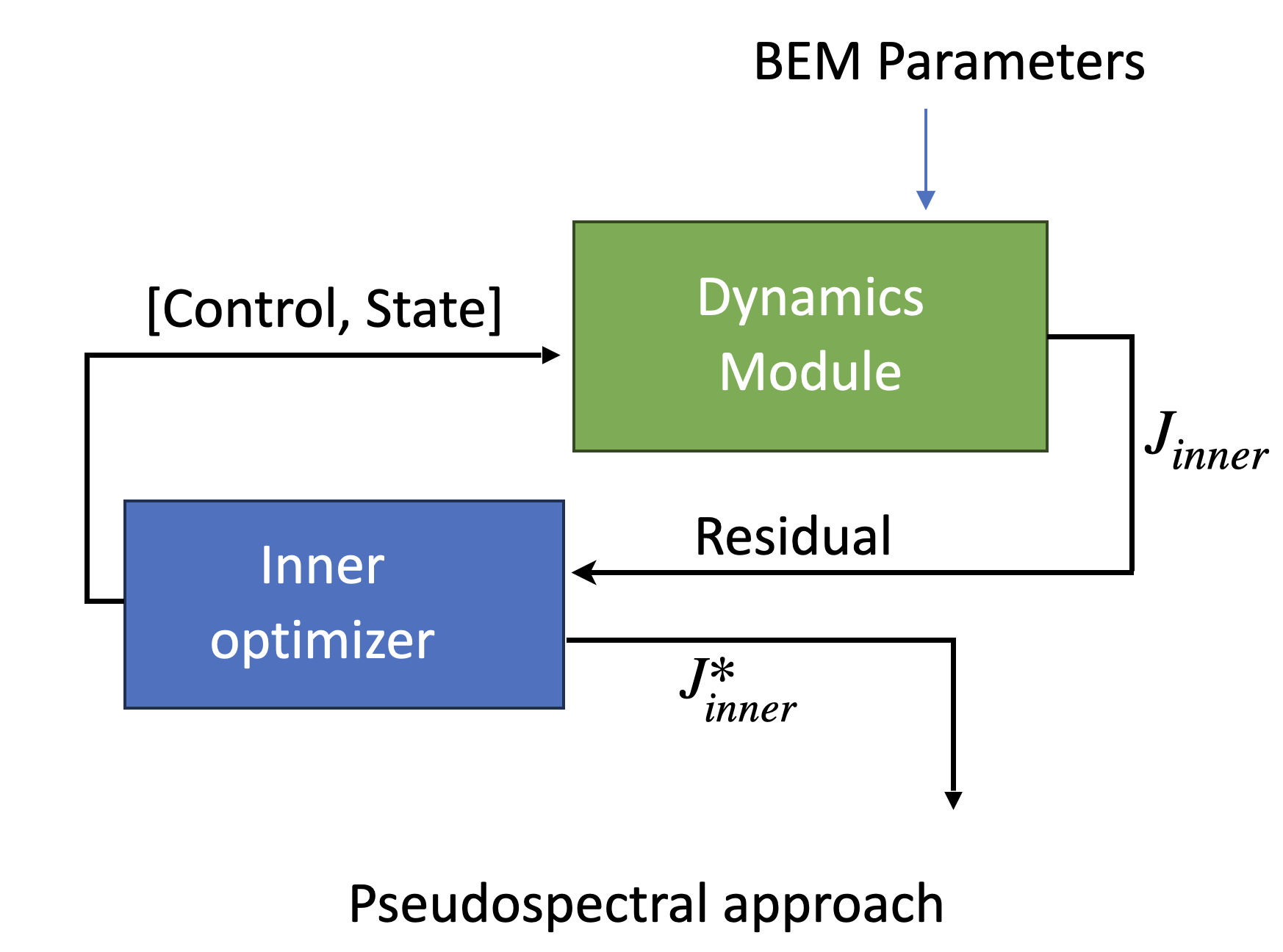}
    \caption{Pseudospectral solver solves for both the state and controls for each WECs}
    \label{fig:ps_ccd}
\end{figure}

As shown in Fig.~\ref{fig:ps_ccd}, the inner pseudo-spectral solver replaces the separate dynamics solver and controller by jointly optimizing the states and control actions for each devices.

In this study, an unstructured controller is employed, representing an idealized ``perfect" model predictive control (MPC) scheme that can predict the future wave excitation with 100\% accuracy. This formulation establishes an upper bound on achievable performance and isolates the effects of geometry and layout on absorbed power. However, unconstrained optimal control often produces infeasible PTO forces, leading to unrealistic WEC motions and power outputs. To ensure physical realization, an inequality constraint is imposed on the PTO force, $f_u < f_{u,\mathrm{max}}$, which is incorporated directly into the control optimization to maintain feasible actuation levels. By embedding this constraint within the optimization rather than applying post hoc saturation, the controller is designed to operate within achievable force limits throughout the motion cycle, ensuring realistic and implementable control behavior.



\subsection{Uncertainty Quantification}
\label{subsec:uq}
Uncertainty in the wave environment introduces variability in WEC farm performance. These aleatoric uncertainties, arising from the stochastic nature of ocean waves, must be propagated through the hydrodynamic and control models to assess system robustness. While a direct Monte Carlo approach could achieve this by sampling the uncertain parameters and computing statistical moments of the model output, it is highly sample‐inefficient, requiring thousands of radiation-diffraction and optimal control evaluations, and is therefore computationally prohibitive and we must look for efficient procedures for the optimization with large number of devices.  

PCE offers a sample-efficient, non-intrusive alternative to traditional Monte Carlo methods by constructing a spectral surrogate that approximates the system response using a small number of deterministic simulations~\citep{pce_review, UQ_review}. In this formulation, the model output is expressed as a sum of orthogonal polynomial basis functions with random coefficients, built using a limited number of samples~\citep{UQ_review}. These basis functions correspond to the underlying probability distributions of the uncertain variables, allowing the surrogate to capture dominant stochastic behavior efficiently. By leveraging these distribution-matched bases, PCE enables accurate estimation of statistical moments with orders of magnitude fewer samples than traditional Monte Carlo methods. Owing to its scalability and computational efficiency, PCE is particularly well suited for high-fidelity MDO problems involving expensive coupled simulations.

In this study, PCE is employed to propagate uncertainty in the significant wave height, $H_s$, peak period, $T_p$, and wave direction, $\beta$, through the coupled hydrodynamic and control models. The power output, $P$, is expressed as a spectral expansion in terms of these orthogonal polynomial bases, and the coefficients are determined via regression. Once constructed, the surrogate directly approximates the system response and quantifies uncertainty through its statistical moments. In this study, the expected absorbed power, $\mathbb{E}[P]$, corresponds to the zeroth-order PCE coefficient ($c_0$) as a consequence of the orthogonality of the polynomial basis. This enables efficient evaluation of system performance under stochastic sea states.

Because the wave parameters are uncertain, the PCE surrogate must be retrained at each design iteration, requiring repeated solution of the diffraction, radiation, and optimal control problems to evaluate the risk-neutral objective. Despite this added complexity, the non-intrusive PCE formulation substantially reduces computational cost while maintaining high fidelity in uncertainty propagation. 

For an uncertain input space defined by random variables $\mathbf{X} = \{X_1, X_2, \dots, X_n\}$, the model output $f(\mathbf{X})$ (in this case, the absorbed electrical power) is approximated by an orthogonal polynomial expansion:

\begin{equation}
    f(\mathbf{X}) \approx \sum_{i=0}^{N} c_i \phi_i(\mathbf{X}),
\end{equation}
where $\{\phi_i(\mathbf{X})\}_{i=0}^{N}$ are orthogonal polynomials corresponding to the joint probability distribution of $\mathbf{X}$ and $\{c_i\}$ are the expansion coefficients. The orthogonal polynomials are constructed to satisfy orthogonality, defined as:
\begin{equation}
    \mathbb{E}\left[\phi_i(\mathbf{X}) \phi_j(\mathbf{X})\right] = 
    \int \phi_i(\mathbf{X}) \phi_j(\mathbf{X}) w(\mathbf{X})\, d\mathbf{X} = \delta_{ij},
\end{equation}
with $w(\mathbf{X})$ representing the joint probability density function of $\mathbf{X}$ and $\delta_{ij}$ is the Kronecker delta. 

The coefficients $c_i$ are obtained via least-squares regression:
\begin{equation}
    \Phi \mathbf{c} = \mathbf{P},
\end{equation}
where $\Phi$ is the matrix of polynomial basis evaluations at training points, and $\mathbf{P}$ is the vector of time-averaged electrical power values computed from deterministic simulations for each sampled realization of the uncertainty. 

The training points are generated using Gaussian quadrature, which leverages Hermite and Legendre polynomials to efficiently sample the uncertain input space. Hermite polynomials are used due to their suitability for capturing the variability of variables following normal distributions, such as the significant wave height, $H_s$, wave period ($T_p$) and wave direction $\beta$ (after Nataf transformation~\citep{UQpy_Nataf}). These orthogonal polynomial bases are selected to correspond to the underlying probability density functions of the random variables, ensuring that the PCE accurately captures the stochastic response of the wave energy converter system.

In this implementation, the PCE surrogate model employs Hermite polynomials that are orthogonal with respect to the standard normal distribution. The PCE basis is constructed for three independent standard normal variables, $(z_{H_s}, z_{T_p}, z_{\beta})$ (after nataf transformation from their empirical distribution), representing the transformed uncertain parameters of significant wave height, period and wave direction. 

The accuracy of the PCE surrogate is assessed using the train-test split of the data collected for random sampling of the parameter. Only the PCE model that has at least $R^2 = 90$ is utilized for the estimation of the mean power for the new realization of the random parameters. 



\section{Optimization Strategy}
\label{sec:wec_array_opt}

The optimization problem is solved using the Covariance Matrix Adaptation Evolution Strategy (CMA-ES), a gradient-free, stochastic, evolutionary algorithm well suited for black-box and potentially noisy continuous optimization problems~\citep{hansen2016cma, pymoo}. In this method, each individual in the population represents a candidate solution, a vector in the design space, with a corresponding fitness value defined by the objective function \( f: \mathbb{R}^d \rightarrow \mathbb{R} \).  

CMA-ES samples new candidates from a multivariate Gaussian distribution,
\begin{equation}
\mathcal{N}(\mu, \sigma_h^2 C),
\end{equation}
where the mean vector $\mu$ directs the search toward promising regions of the design space, the step size $\sigma_h$ controls the overall exploration scale, and the covariance matrix $C$ shapes the orientation of the search. Both $\mu$ and $C$ are iteratively updated across generations based on the fitness of sampled solutions, enabling the algorithm to adaptively learn the problem’s landscape and progressively concentrate search efforts in high-performing regions~\citep{Hansen2006}. CMA-ES was selected for this study due to its proven effectiveness in black-box and potentially noisy continuous optimization problems~\citep{Hansen2006} and its robustness to non-smooth objective landscapes. The variation in the objective evaluations may come from the randomness in the PCE-based training process, specifically the sampling of data and the training/test split used to build the regression model that approximates the output described in Section \ref{subsec:uq}

The CMA-ES optimizer was initialized with the scaled design vector \(x_0\) and a step-size parameter of \(\sigma = 0.25\). Variable bounds were enforced as \((x_\mathrm{l},\, x_\mathrm{u})\), and termination tolerances were set to \(\mathrm{tol}_x = 10^{-3}\) and \(\mathrm{tol}_f = 10^{-3}\), corresponding to convergence when either the change in design variables or the objective value falls within the order of \(1\,\mathrm{W}\). The algorithm was limited to a maximum of 198 iterations, with a population size of 50 individuals per generation.  

The full optimization process follows a nested, or ``double-loop," structure in which the optimizer (outer loop) proposes design vectors, while the multidisciplinary simulation and PCE modules (inner loop) evaluate the expected objective. This coupling between WEC power optimization and PCE training across independent wave parameter distributions substantially increases computational cost. To maintain numerical efficiency, a separate PCE surrogate for uncertainty calculation is constructed at each design iteration. Each surrogate then predicts the expected power response for a given design vector $\mathbf{x}$ across the stochastic space defined by $(H_s, T_p, \beta)$. The complete procedure is summarized in Algorithm~\ref{alg:pce_optimization}. By integrating surrogate-based uncertainty quantification within the MDO architecture, the approach achieves a balance between computational efficiency and robust design fidelity.

\begin{figure}[t!]
    \centering
    \includegraphics[width=\linewidth]{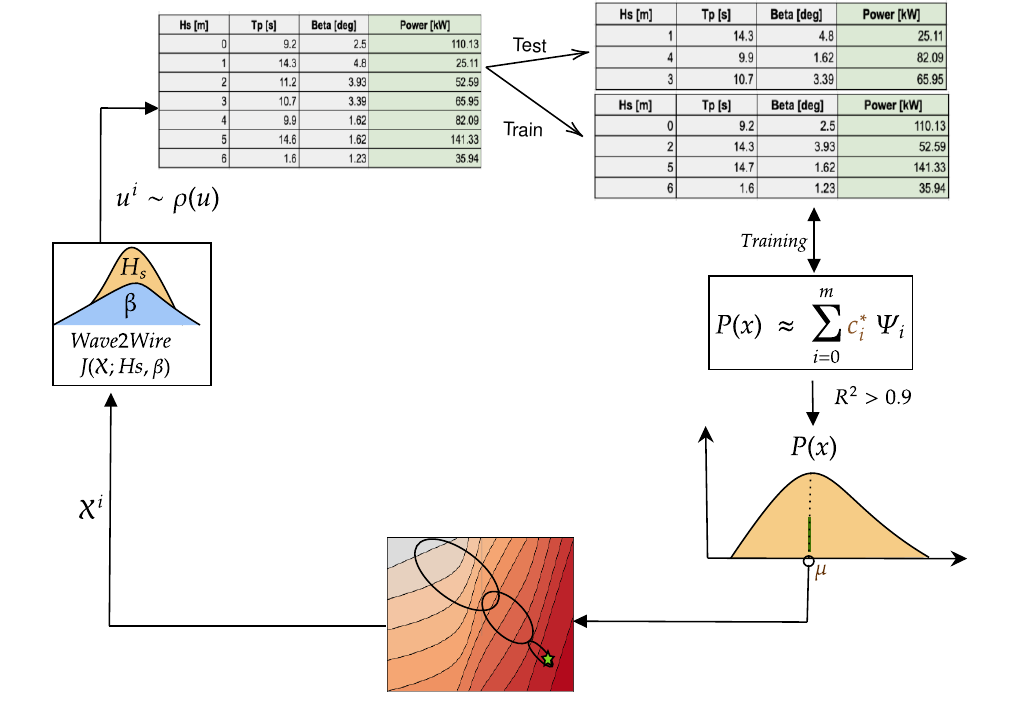}
    \caption{Workflow of the MDO framework integrating PCE-based uncertainty quantification. At each design iteration, deterministic simulations are performed to train the PCE surrogate, which approximates the expected power response and passes it to the optimizer.}
    \label{fig:pcegraphics}
\end{figure}

\small{
\begin{algorithm}[H]
\caption{Stochastic Design Optimization}
\label{alg:pce_optimization}

\textbf{Input:} Initial design variables \(x_0\) (e.g., WEC radii, heights, positions), wave parameter distributions \(\mathbf{u}\), maximum allowable samples \(N_{\text{max}}\), accuracy threshold for surrogate \(R^2_{\text{min}}\) (e.g., 0.90).\\
\textbf{Output:} Optimized design variables \(x^*\), expected power output \(\mathbb{E}[P(x^*)]\).

\begin{enumerate}
    \item \textbf{Outer Optimization Loop:}
    \begin{enumerate}
        \item Initialize design variables \(x = x_0\).
        \item Repeat until convergence:
        \begin{enumerate}
            \item \textbf{Inner PCE Training for Given \(x\):}
            \begin{enumerate}
                \item \textbf{Sampling:} Generate \(N = 20\) independent samples of wave parameters \(\mathbf{u}\).
                \item \textbf{Model Construction:} Train a Polynomial Chaos Expansion (PCE) model to approximate \(P(x)\):
                \[
                P(x) \approx \sum_{i=0}^m c_i \Psi_i(\mathbf{u}),
                \]
                where \(c_i\) are the coefficients and \(\Psi_i(\mathbf{u})\) are the basis functions corresponding to the distribution of random variables.
                \item \textbf{Model Validation:} Perform a train-test split and compute the \(R^2\)-score on the test dataset.
                \begin{itemize}
                    \item If \(R^2 \geq R^2_{\text{min}}\), use the model for optimization.
                    \item If \(R^2 < R^2_{\text{min}}\), proceed to Step (d).
                \end{itemize}
                \item \textbf{Adaptive Sampling:} Incrementally add more samples (N = N+10) to the training and test datasets to improve accuracy.
                \begin{itemize}
                    \item If \(N \geq 40\), switch to statistical averaging of the outputs.
                \end{itemize}
            \end{enumerate}
            \item \textbf{Design Evaluation:}
            \begin{enumerate}
                \item Use the validated PCE model to compute the expected power output \(\mathbb{E}[P(x)]\).
                \item Update design variables \(x\) based on the optimization algorithm (using CMA-ES optimizer).
            \end{enumerate}
        \end{enumerate}
    \end{enumerate}
    \item \textbf{Termination:} Stop when the design variables \(x\) converge, or a predefined maximum number of iterations is reached.
\end{enumerate}
\end{algorithm}
}
The above algorithm can be visualized as a flowchart as shown in Figure ~\ref{fig:pcegraphics} that shows the deterministic simulations for realization of random parameters and the PCE model training to approximate the output distribution.

\section{Notice}
Results, discussion, conclusions, and future work sections are forthcoming pending the results of the optimization experiments.

All analysis scripts, optimization routines, and case studies used in this work are open-source and available at:  
\url{https://github.com/symbiotic-engineering/wec_array_opt}. This repository supports reproducibility and provides a foundation for further research on robust control co-design and multidisciplinary optimization of marine energy systems.



\section{CRediT authorship contribution statement}
\textbf{Kapil Khanal and Nate DeGoede:} Conceptualization, Methodology, Software, Formal analysis, Investigation, Data Curation, Original Draft, Review \& Editing, Visualization. \textbf{Dr. Maha N Haji:} Resources, Supervision, Writing - Review \& Editing, Funding acquisition

\section{Acknowledgments} 
The authors would like to thank Olivia Vitale (for extensive reviews),
Carlos S. Michelen, Matthieu Ancellin, Rebecca McCabe and Matthew Haefner for their valuable feedback on the simulation and analysis in this manuscript. This work was supported in part by the Graduate Fellowship (for first year doctoral students) via Systems Engineering Department.

\bibliographystyle{elsarticle-harv} 
 \bibliography{references}


\end{document}